\documentclass[12pt]{article}

\pdfoutput=1

\usepackage[pdftex]{graphicx}
\usepackage{amssymb}

\usepackage{bbm}

\usepackage{color}

\newcommand{\rf}[1]{(\ref{#1})}
\newcommand{\beq}{\begin{equation}}
\newcommand{\eeq}{\end{equation}}
\newcommand{\bea}{\begin{eqnarray}}
\newcommand{\eea}{\end{eqnarray}}

\newcommand{\e}{\mbox{e}}

\newcommand{\lam}{\lambda}

\renewcommand{\a}{\alpha}

\newcommand{\del}{\delta}

\newcommand{\oh}{\frac{1}{2}}

\newcommand{\dg}{\dagger}

\newcommand{\tr}{\mathrm{Tr}\,}
\newcommand{\ra}{\rangle}
\newcommand{\la}{\langle}
\newcommand{\prt}{\partial}

\newcommand{\tH}{{\tilde{H}}}

\newcommand{\hH}{{\hat{H}}}

\newcommand{\vac}{|0\ra}

\newcommand{\vaccu}{\langle{\rm vac}|}
\newcommand{\cuum}{|{\rm vac}\rangle}

\newcommand{\Wtre}{W^{(3)}}

\begin{document}

\begin{center}
\vspace{24pt}
{ \large \bf
  Creating 3, 4, 6 and 10-dimensional spacetime
\\
  from W3 symmetry}

\vspace{24pt}

{\sl J.\ Ambj\o rn}$\,^{a,b}$,
and {\sl Y.\ Watabiki}$\,^{c}$

\vspace{10pt}

{\small

$^a$~The Niels Bohr Institute, Copenhagen University\\
Blegdamsvej 17, DK-2100 Copenhagen \O , Denmark.\\
email: ambjorn@nbi.dk
\vspace{10pt}

$^b$~Institute for Mathematics, Astrophysics and Particle Physics
(IMAPP)\\ Radbaud University Nijmegen, Heyendaalseweg 135, 6525 AJ, \\
Nijmegen, The Netherlands

\vspace{10pt}

$^c$~Tokyo Institute of Technology,\\ 
Dept. of Physics, High Energy Theory Group,\\ 
2-12-1 Oh-okayama, Meguro-ku, Tokyo 152-8551, Japan\\
{email: watabiki@th.phys.titech.ac.jp}

}

\end{center}

\vspace{24pt}

\begin{center}
{\bf Abstract}
\end{center}

We describe a model where breaking of W3 symmetry will
lead to the emergence of time and subsequently of space.
Surprisingly the simplest such models which lead to higher 
dimensional spacetimes are based on the four ``magical'' Jordan 
algebras of 3x3 Hermitian  matrices 
with  real, complex, quaternion and octonion entries, respectively. 
The simplest symmetry breaking leads to universes 
with spacetime dimensions 3, 4, 6, and 10.

%\vfill

\newpage

\section{Introduction}\label{intro}

String field theory is notoriously complicated, but there is a baby version,
namely non-critical string field theory \cite{NCSFT,watabiki,aw}.  
Non-critical string theory describes two-dimensional quantum gravity 
coupled to a conformal field theory with a central charge $c < 1$ 
and the corresponding string field theory aims 
to describe the dynamics of merging and splitting of such strings. 
For $c=0$ the situation is particularly simple. 
One has creation and annihilation operators $\Psi^\dg(L)$ and 
$\Psi(L)$ for spatial universes of length $L$. 
An even simpler string field theory exists, 
CDT string field theory \cite{GSFT}. 
The starting point is the continuum limit of 2d 
``causal dynamical triangulations" (CDT) \cite{al,GCDT}, 
a limit which in the case of trivial spacetime topology 
is 2d quantum Horava-Lifshitz gravity \cite{agsw}
(for higher dimensional CDT which might also be related 
to Horava-Lifshitz gravity \cite{horava}, 
see e.g. \cite{higherCDT,physrep}). 
CDT string field theory describes 
the dynamics of topology changes of spacetime.
%, starting from 2d quantum Horava-Lifshitz gravity. 
The Hamiltonian involves terms like
\beq\label{zj1}
\Psi^\dg(L_1)\Psi^\dg(L_2)\Psi(L_1+L_2),\quad\quad
\Psi^\dg(L_1+L_2)\Psi(L_2)\Psi(L_1),
\eeq
which describe the annihilation of a  universe of length $L_1+L_2$ and the 
creation of two universes of lengths $L_1$ and $L_2$, or the reverse process
\cite{GSFT}. The interaction term in the string field Hamiltonian thus contains 
products of three annihilation and creation operators. The same is true
for standard non-critical string theory which is related to a special kind
of $\Wtre$ symmetry which ensures that the partition function can be viewed
as a $\tau$-function of certain coupling constants \cite{kdv}. 
This led us to realize that 
one can obtain the CDT string field theory starting 
from a so-called $\Wtre$ Hamiltonian by symmetry breaking \cite{awBB}. 
The $\Wtre$ Hamiltonian has a natural, so-called 
{\it absolute} vacuum and offers no obvious spacetime interpretation, but 
breaking the $\Wtre$-symmetry led to a so-called {\it physical} vacuum and 
the emergence of time and  one spatial dimension. 
The purpose of this article is to generalize the construction 
such that one can create universes 
with one time direction and higher dimensional spaces. 
The simplest symmetry breaking leads to spacetime dimensions 
2+1, 3+1, 5+1 and 9+1. 

In Sec.\ \ref{reviewW3} we shortly review the $\Wtre$ Hamiltonian formalism 
introduced in \cite{awBB} and in Sec. \ref{jordan} 
we generalize it to include internal degrees of freedom. 
This leads to the introduction of the so-called magical Jordan algebras 
and it is the structures of these algebras 
which result in spacetime dimensions 2+1, 3+1, 5+1 and 9+1.

\section{The $\Wtre$ Hamiltonian}\label{reviewW3}

The formal definition of a $W^{(3)}$ algebra in terms of 
operators $\a_n$ satisfying   
\beq\label{jx10} 
[\a_m,\a_n] = m \,\del_{m+n,0}.
\eeq
is the following
\beq\label{jx11}
\a(z) = \sum_{n \in \mathbbm{Z}} \frac{\a_n}{z^{n+1}},
\qquad
W^{(3)}(z) = \frac{1}{3}\! :\!\a(z)^3\!:
\;= \sum_{n\in Z} \frac{W^{(3)}_n}{z^{n+3}}.
\eeq
The normal ordering $:\!\!(\cdot)\!\!:$ 
refers to the $\a_n$ operators ($\a_n$ to the 
left of $\a_m$ for $n>m$)\footnote{We remark that this ordering is 
opposite to the standard ordering one would use in conformal field
theory. One can obtain the conventional ordering by the so-called
$\star$-operation \cite{aw}.} 
and we have
\beq\label{jx12}
W^{(3)}_n = \frac{1}{3} \sum_{k,l,m} :\!\a_k\a_l\a_m\!: \del_{k+l+m,n}.
\eeq 
We then define the ``absolute vacuum'' $|0\ra$ by the following condition:
\begin{equation}\label{jy2}
\a_n |0\rangle =  0,\quad n < 0.
\end{equation}
and the so-called $W$-Hamiltonian ${\hH}_{\rm W}$ by 
\beq\label{jy3}
{\hH}_{\rm W} := - W^{(3)}_{-2} =
-\,\frac{1}{3} \sum_{k, l, m}
:\!\alpha_k \alpha_l \alpha_m\!: \delta_{k+l+m,-2}.
\eeq
Note that ${\hH}_{\rm W}$ does not contain any coupling constants.

It was  shown in \cite{awBB} that by introducing a coherent state, 
which is an eigenstate of $\a_{-1}$ and $\a_{-3}$ and which 
we denoted the ``physical'' vacuum state $\cuum$, ${\hH}_{\rm W}$ was 
closely related to the CDT string field Hamiltonian   $\hH$. 
We thus defined 
\beq\label{zj5}
\cuum \propto \e^{ \lam_1 \a_{1} + \lam_3 \a_3} \vac,
\eeq
and we have 
\beq\label{zj6}
\a_{-1} \cuum =\lam_1 \cuum, \qquad \a_{-3} \cuum= 3 \lam_3 \cuum.
\eeq
The main point is the  following: because $\vaccu \a_n \cuum$
is different from zero for $n=-1$ and $n=-3$, 
${\hH}_{\rm W}$ will now contain terms only involving two operators $\a_l$. 
These terms can act like quadratic terms in $\hH$. 
At the same time the cubic terms left in ${\hH}_{\rm W}$ 
will act like the interaction terms in $\hH$, 
resulting in joining and splitting of universes. 
Finally, the expectation values of $\a_{-1}$ and $\a_{-3}$ 
determine the coupling constants of $\hH$.
More precisely one has \cite{awBB}
 \beq\label{zj2}
 {\hH}_{\rm W}  \propto \hH + c_4 \phi_4^\dg +c_2 \phi^\dg_2
 %+ c_1 \phi_1^\dg + c_0,
 \eeq
where $\hH$ is the CDT string field Hamiltonian. 
$c_4$ and $c_2$ are constants.
%determined by $\lambda_1$ and $\lambda_3$ uniquely.
The creation operators $\phi^\dg_n$ are the $\a_n$, $n > 0$, 
while annihilation operators $\phi_n$ are 
related to $\a_n$, $n < 0$, except that 
$\phi_1$ and $\phi_3$ are shifted by eigenvalues given in eq.\ \rf{zj6}, 
such that $\phi_n \cuum =0$. 
$\hH$ is normal ordered such that $\hH \cuum = 0$.
 
By breaking the $\Wtre$ symmetry one can thus obtain CDT string field theory
except for one important point: the vacuum is not stable. 
The terms  $c_4 \phi_4^\dg +c_2 \phi^\dg_2$ %$+ c_1 \phi_1^\dg $ 
cause universes of infinitesimal length to be created 
and the non-interacting part of $\hH$, 
which explicitly can be written as 
\beq\label{zj3}
 \hH_{0} =- \sum_{l=1}^\infty \phi_{l+1}^\dagger l \phi_l
+ \mu \sum_{l=2}^\infty \phi_{l-1}^\dagger l \phi_l,
\eeq
might expand such an infinitesimal length  space to macroscopic size. 
The relation between the operators $\phi_l,~\phi^\dg_l$ and the operators 
$\Psi(L),~\Psi^\dg(L)$ which  annihilate and create spatial universes of 
macroscopic length $L$ is as follows 
\beq\label{jj1}
\Psi^\dagger(L) = \sum_{l=0} \frac{L^l}{l!} \; \phi^\dg_l
%,\qquad
%\phi_l = \int_0^\infty \d L \;
%\frac{L^l}{l!} \; \Psi(L)
.
\eeq
When expressed in terms of $\Psi(L)$ and $\Psi^\dg(L)$ 
the Hamiltonian \rf{zj3} can be written as 
\beq\label{j6}
\hH_0 = \int_0^\infty dL \; \Psi^\dg (L) 
H_0  \Psi(L),\quad\quad H_0= -  L\frac{\prt^2}{\prt L^2}+\mu L,
\eeq
where the two terms on the rhs of eq.\ \rf{j6} corresponds to the two terms 
on the rhs of eq.\ \rf{zj3}. It should now be clear why we denote the two 
terms on the rhs of eq.\ \rf{zj3} the kinetic and the cosmological term, respectively.
$H_0$ is  the original CDT Hamiltonian \cite{al} for the evolution of a single
2d universe, without topology changes. If  $\mu > 0$ a universe
starting with zero (or more precisely infinitesimal) length will have a 
(unnormalized) wave function and a corresponding expectation value
of the size of space at time $T$:
\beq\label{yj1}
G(L,T) = \mu L \; 
\frac{\e^{-\sqrt{\mu} L \coth (\sqrt{\mu}\,T)}}{\sinh^2 (\sqrt{\mu}\,T)} 
\quad\quad
\la L \ra = \frac{1}{\sqrt{\mu} }\tanh (\sqrt{\mu}\,T).
\eeq 
If $\mu < 0$ the corresponding equations become ($ \tilde{\mu} = - \mu$):
\beq\label{yj2}
G(L,T) = \tilde{\mu} L \; 
\frac{\e^{-\sqrt{\tilde{\mu}} L \cot (\sqrt{\tilde{\mu}}\,T)}}
     {\sin^2 (\sqrt{\tilde{\mu}}\,T)} 
\quad\quad
\la L \ra = \frac{1}{\sqrt{\tilde{\mu}}} \tan (\sqrt{\tilde{\mu}}\,T).
\eeq 
In this case 
the wave function only  belongs to the Hilbert space of $H_0$ for 
$0 < T < \pi/(2\sqrt{\tilde{\mu}})$. 
At $T =  \pi/(2\sqrt{\tilde{\mu}})$ 
the universe has expansed to infinite size.

\section{Generalization to higher dimensions}\label{jordan}

The above creation of space and time, described in \cite{awBB},
is limited to one space and one time dimension. In order  to create $d$-dimensional
space we introduce an internal index $a$, $a=1,\ldots,d$ and consider the corresponding
extended $\Wtre$ algebra. The classification  of such $\Wtre$ algebras is closely related 
to the classification of Jordan algebras (see \cite{romans} for a review of $W$ algebras 
and their relations to Jordan algebras) and surprisingly it turns out that only 
the four so-called magical Jordan algebras allow us to make  symmetry 
breakings which lead to  CDT-like Hamiltonians of the kind considered above.
We will discuss the reason for that elsewhere \cite{to-come}, and here we will
just  review how one defines the four magical Jordan algebras and the corresponding 
$\Wtre$ Hamiltonians.

Let $H_3(\mathbbm{F})$ denote the $3\times 3$  Hermitian 
matrices with entries in $\mathbbm{F}$, where $\mathbbm{F} = 
\mathbbm{R}, \mathbbm{C}, \mathbbm{H}$ and $\mathbbm{O}$ (the real numbers,
the complex numbers, the quaternions and the octonions). 
The $H_3(\mathbbm{F})$'s are real vector spaces of dimensions 6, 9, 15 and 27, 
and they are Jordan algebras 
when one defines the algebra multiplication of two elements 
as the anti-commutator of the corresponding matrices:
\beq\label{jordan1}
 X \circ Y := \oh \{X,Y\}.
 \eeq
 If one defines the scalar product on $H_3(\mathbbm{F})$ by
 \beq\label{jordan1a}
 \la X,Y\ra = \oh \,\tr (X \circ Y),
 \eeq
 it has  an orthogonal decomposition
 \beq\label{jordan1b}
 H_3(\mathbbm{F}) = \mathbbm{R}\cdot {1}_{3\times 3}  \oplus 
 \tH_3(\mathbbm{F}),
\eeq
where $\tH_3(\mathbbm{F})$ denotes the traceless matrices. 
The $\tH_3(\mathbbm{F})$'s are 
real vector spaces of dimensions $d=5$, 8, 14, 26, respectively.
Let $E_a$ [\,$a=1,\ldots,d$\,] denote an orthonormal basis 
of the vector space $\tH_3(\mathbbm{F})$. 
The structure  constant of the Jordan algebra can be defined as 
\beq\label{jordan2}
 d_{abc} := \oh \,\tr ((E_a\circ E_b) \circ E_c).
\eeq
The structure constants are invariant 
under the action of the automorphism groups of the algebras, 
which are $SO(3)$, $SU(3)$, {\it U}$Sp(6)$ and $F_4$ respectively.
In the case of $\tH_3(\mathbbm{C})$ 
(equal as a real vector space to the Lie algebra $su(3)$) 
we can choose as the $E_a$ 
the standard Gell-Mann matrices $\lam_a$ of $su(3)$, 
which satisfy the anti-commutation relation
\beq\label{jordan3}
\{ \lam_a,\lam_b\} = \frac{4}{3}\del_{ab}\cdot {1}_{3\times 3} + \frac{1}{2} \sum_c d_{abc} \lam_c 
\eeq
$\lambda_a^{R}$, $a=1,3,4,6,8$ are 5 real, symmetric matrices 
and they form the 5-dimensional basis $E_a$ for $\tH_3(\mathbbm{R})$. 
$A_b = -i \lambda_b$, $b=2,5,7$ are real antisymmetric matrices. 
When multiplied by $i$ they form together with the $\lambda^{R}_a$ matrices 
the 8-dimensional basis $E_a$ for $\tH_3(\mathbbm{C})$, as already mentioned. 
If the $A_b$ matrices are multiplied  by $i,j,k$, 
the generalised imaginary quaternions numbers 
they form together with the 5 $\lambda^{R}_a$ 
the 14-dimensional basis $E_a$ of $\tH_3(\mathbbm{H})$.
Finally, 
if the $A_b$ matrices are multiplied by the 7 generalised imaginary 
octonion numbers $i,j,k,\bar{i},\bar{j},\bar{k},\ell$ 
they form together with the $\lambda^{R}_a$ matrices 
the 26-dimensional basis $E_a$ for $\tH_3(\mathbbm{O})$. 

The generalization of \rf{jx10} - \rf{jy3} is now straightforward. 
We define the current
\beq\label{hj1}
\a(z) = \sum_a \a^{(a)}(z)\,E_a,
\qquad
\a^{(a)}(z) = \sum_{n\in \mathbbm{Z}} \frac{\a_n^{(a)}}{z^{n+1}},
\eeq
\beq\label{hj2}
W^{(3)}(z) = \frac{1}{3} :\! \tr ((\a(z) \circ \a(z) ) \circ \a(z)) \!:
\:= 
\sum_{n\in \mathbbm{Z}} \frac{W^{(3)}_n}{z^{n+3}},
\eeq
where the commutation relations are
\beq\label{hj10} 
[\a_m^{(a)},\a_n^{(b)}] = m \,\del_{m+n,0} \,\delta_{a,b},
\eeq
and  we find for the 
$\Wtre$ Hamiltonian the expression
\beq\label{jordan4}
{\hH}_{\rm W} := -W_{-2}^{(3)} =
-\,\frac{1}{3} \sum_{k, l, m} \sum_{a, b, c} d_{abc}
:\!\alpha_k^{(a)} \alpha_l^{(b)} \alpha_m^{(c)}\!: \delta_{k+l+m,-2}.
\eeq
Again, this model only allows a spacetime interpretation 
after choosing a specific coherent state. 
There are some interesting choices, 
but here we will only discuss the simplest ones, 
namely some choices of  breaking 
in the 8-direction and the 3-direction. 
Instead of \rf{zj5} and \rf{zj6} we can choose 
\beq\label{zj5b}
\cuum_{8} \propto
\e^{ \lam_1^{(8)} \a_{1}^{(8)} + \lam_3^{(8)} \a_3^{(8)}}\vac,
\eeq
and we have 
\beq\label{yj6}
\a_{-1}^{(8)}  \cuum_{8} =\lam_1^{(8)}  \cuum_{8},
\qquad
\a_{-3}^{(8)}  \cuum_{8} = 3 \lam_3^{(8)}  \cuum_{8}.
\eeq

When one looks at the coefficients $d_{ab8}$ 
in order to obtain the non-interacting part of the Hamiltonian 
(the equivalent of $\hH_0$ given by \rf{zj3}), 
firstly it is observed that 
only coefficients $d_{aa8}$ are different from zero. 
This implies that the non-interacting part of the Hamiltonian is diagonal 
in the ``space'' indices $a$. Next, the only coefficient $d_{aaa}$ 
which is different from zero is $d_{888}$. Thus  the only $\phi^{(a)}$ field 
which has a cubic self-interaction is the 8-field. 
The $d_{aa8}$ have the following values 
\beq\label{zj7}
d_{aa8} = \frac{1}{\sqrt{3}}
\ \ 
\hbox{or}
\ 
-\frac{1}{2\sqrt{3}},
\qquad
d_{888} = -\,\frac{1}{\sqrt{3}}.
\eeq
If we use the vacuum $ \cuum_{8}$, 
the two groups will result in Hamiltonians with opposite signs. 
Let us assume that
the symmetry breaking is chosen such that the non-interacting 
Hamiltonians with $d_{aa8} > 0$ will
correspond to the CDT Hamiltonians of the form \rf{j6}, 
only carrying now an index $a$. 
The Hamiltonians with negative $d_{aa8}$ 
will now have a negative kinetic term. 
We have several options when addressing the negative Hamiltonians. 
The state $| l \ra^{(a)} = (\phi_l^{(a)})^\dg \cuum$ has macroscopic length $L =0$. 
This wave function is basically $l$-times the derivative of $ \delta (L)$. 
In the case where the internal index $a$ is such that the kinetic part of  the  Hamiltonian 
$\hH^{(a)}_0$ is positive, the time evolution of such an initial state, 
created from the vacuum by the creation operators in \rf{zj2} (with internal index $a$),
will have the time evolution shown in eqs.\ \rf{yj1} or \rf{yj2} 
(ignoring the cubic interaction terms in the Hamiltonian). 
In some time interval the wave functions thus belong 
to the Hilbert space of Hamiltonian. 
However, if the internal index $a$ is such that the kinetic term of $\hH^{(a)}_0$ 
is negative, the time evolution does not give us an acceptable wave function 
(it is obtained from \rf{yj1} by changing the sign of $T$). 
Thus we can choose to insist that a macroscopic state with 
macroscopic length is never created  in this way for modes where $d_{aa8}$ 
is negative. 
Let us first accept this viewpoint and assume that macroscopic directions 
with $d_{aa8} < 0 $ are not excited 
by acting with $\phi^{(a)\dg}_l$ on $\vac_8$. 

When we then look at the four magical algebras, we have 
for $\tH_3(\mathbbm{R})$ two $a$ where $d_{aa8} = 1/\sqrt{3}$. 
For $\tH_3(\mathbbm{C})$ we have three $a$'s where $d_{aa8} = 1/\sqrt{3}$.
For $\tH_3(\mathbbm{H})$ we have five $a$'s where $d_{aa8} = 1/\sqrt{3}$ 
and finally for $\tH_3(\mathbbm{O})$ 
we have nine $a$'s where $d_{aa8} = 1/\sqrt{3}$. 
The symmetry breaking 
corresponds to breaking the automorphism group of the $\Wtre$ algebras 
from $SO(3)$ to $SO(2)$, 
from $SU(3)$ to $SO(3)$, 
from {\it U}$Sp(6)$ to $SO(5)$ and
finally from $F_4$ to $SO(9)$. 
The extended spacetime dimensions will be (including the time) 
2+1, 3+1, 5+1 and 9+1 
which are the dimensions of the classical superstrings 
(the quantum superstring only survives in the ten-dimensional case).

However, it is possible to take another point of view. 
If we consider the modes $\a_l$
or $\phi_l$ as the fundamental variables, 
it is up to us to find a continuum length 
interpretation. For the positive Hamiltonian \rf{jj1} was fine. 
If we obtain a negative 
Hamiltonian with this prescription, we are free to change \rf{jj1} to 
\beq\label{jk1}
\Psi^\dagger(L) = \sum_{l=0} \frac{(-L)^l}{l!} \; \phi^\dg_l
%,\quad \phi_l = \int_0^\infty \d L \;
%\frac{(-L)^l}{l!} \; \Psi(L)
.
\eeq
This will simple change the sign of $L$ and thus the sign of the Hamiltonian
since it is  linear in $L$. With such a change for the $a$'s where the 
kinetic part of $\hH$ is negative,
all directions now have a positive kinetic Hamiltonian and all directions now
have the potential to develop a macroscopic length. However, there will not 
isotropy between the directions anymore since the values of $|d_{aa8}|$ will
fall in three groups so the global symmetry in space will be more complicated.

Let us finally mention that we can refine the symmetry breaking by 
multiplying \rf{zj5b} by a coherent state operator in the 3-direction
\beq\label{zj9}
\cuum_{8,3} \propto \e^{ \lam_1^{(3)} \a_{1}^{(3)}}\cuum_{8}.
\eeq
With this choice, the kinetic terms will be as before, positive and negative.
If we set the $L$ assignment mentioned above such that all kinetic terms 
become positive, the mass matrix takes a form such that we have 
2, 3, 5 or 9 space directions which follow eq.\ \rf{yj2}, 
i.e.\ expand to infinity, 
and the rest of the space directions, 2, 4, 8 or 16, 
stay bounded by a fixed radius,  like in eq.\ \rf{yj1}.
Especially, in the case of $\tH_3(\mathbbm{O})$,
the number of space directions expanding to infinity is 9 and 
the number of space directions staying compact is 16. 
The reader might recall that these numbers 
of space dimensions (non-compact and compact) 
are also encountered in the case of the heterotic string. 
Thus, by imposing an additional symmetry breaking 
in the 3-direction of the internal space 
we have obtained  symmetry breaking patterns 
which resemble the first ones discussed, 
where we simply disregarded directions with negative kinetic terms. 
A more detailed analysis of the symmetry breaking 
pattern will appear elsewhere \cite{to-come}.

\section{Discussion}

The starting point for introducing the above mentioned model 
was that even the baby versions of string field theory, 
the non-critical string field theories,
tell us surprising little about the actual creation of the universe. 
We wanted a starting point from which space and time could emerge, 
maybe (illustrating our lack of creativity) by some symmetry breaking. 
Such a toy model was presented in \cite{awBB} 
and by construction it was a 1+1 dimensional model.
We have now tried to generalize this approach  to higher dimensions, 
to create some dimension enhancement mechanism, 
somewhat inspired by string theory 
where the dimension of spacetime 
is ``just" given by the number of Gaussian fields $X^a$ 
which appears in the string action. 
Since our starting point was a $\Wtre$ algebra, 
we were naturally led to $\Wtre$ algebras with intrinsic symmetry. 
These are related to Jordan algebras and 
to our surprise, only the four so-called magical Jordan algebras 
seemed to lead to a simple generalization of the one-component model 
studied in \cite{awBB}.
Of course a large number of questions need to be clarified.
Let us discuss a number of the issues.
%Of course a large number of questions need to be clarified before our 
%model can be viewed as really creating the kind of continuous spacetime 
%we would like it to create. 
%Let us discuss a number of the issues. 

Firstly, we have not at all discussed any dynamical mechanism 
leading to the symmetry breaking of our $\Wtre$ model. 
If it is spontaneous, 
we have not yet found any {\it natural} mechanism 
which would lead to such symmetry breaking. 
%Also one should be open minded about other mechanisme 
%not necessarily based on spontaneously symmetry breaking. 
%However, if it is {\it not} spontaneous, one encounters a different situation.
%In order to realize the emergence of time,
%``spontaneous" should be thrown away
%because the both concepts contradict.
%One possibility of the symmetry breaking without ``spontaneous" 
%is realized by the interaction
However, one can imagine other ways of realizing the symmetry breaking.
It is possible to have an interaction such that 
\beq\label{intCDT}
\cuum_{\lambda_3,\lambda_1}
\ \longleftrightarrow \ 
\cuum_{\lambda_3',\lambda_1'}
\otimes
\cuum_{\lambda_3'',\lambda_1''}.
\eeq
Details of how to implement this will be published elsewhere \cite{to-come}.
If we denote the theory described by $\hH_{\rm W}$  as a ``first quantized theory",
it would represent a ``second quantization", in the sense that we then introduce 
a quantum theory for the $\lambda$'s, which were before just c-numbers which 
labeled the different ``vacua'' and thus different coupling constants of the universes
created.  
%which is essentially equivalent to the interactions 
%in the critical string field theory.
%This interaction is the second quantization
%if the theory described by $\hH_{\rm W}$ is the first quantization.
%So, if the theory which desribes 
%the creation and the annihilation of universes 
%is the third quantization, 
%the theory described by the interactions \rf{intCDT} 
%will be called the ``fourth quantization". 
Via  an  interaction which allows the process \rf{intCDT}, 
the absolute vacuum $|0\rangle = \cuum_{0,0}$
can become a physical vacuum $\cuum_{\lambda_3,\lambda_1}$ 
and then  time will emerge, as described in \cite{awBB}. After the emergence of time
$\hH_{\rm W}$ can trigger the creation of macroscopic space.

Secondly, how should we really think about the $\Wtre$ symmetry? 
The four magical Jordan algebras lead to four classical $\Wtre$ symmetries. 
However, it is not easy 
to promote these symmetries to quantum symmetries \cite{quantumW}.
The commutators $[W^{(3)}_m,W^{(3)}_n]$ may lead to $W^{(4)}$ operators. 
For a number of $\Wtre$ algebras these $W^{(4)}$ operators 
can be rewritten in a consistent way as a product of $W^{(2)}$ operators 
(i.e.\ Virasoro algebra operators),
and we have a closed $W^{(2)}$, $W^{(3)}$ algebra 
realized via the free bosonic currents $\a (z)$ defined as in \rf{jx11}. 
However, for the four magical algebras this does not work. 
These  algebras then have to be viewed as embedded in 
some larger algebras. Whether it should be the $W^{(\infty)} $ algebra, 
which contains all higher spin components $W^{(N)}$, 
or, as suggested in \cite{hull},
one should use a different decomposition, is not known. 
Following the line of thinking in \cite{hull}, 
there {\it seems} to be an interesting algebraic structure related to 
the magical Jordan algebras, even at the quantum level. This is due to some 
interesting algebraic properties of the structure constants $d_{abc}$ for the
magical Jordan algebras. Naively, the extended W algebra consists of
\bea
&&
W^{(2|\alpha,\beta)}(z) =
\frac{1}{2}
\sum_{a,b} \delta_{ab}
 : \partial^\alpha \alpha^{(a)}(z) \partial^\beta \alpha^{(b)}(z) :,
\nonumber\\
&&
W^{(3|\alpha,\beta,\gamma)}(z) =
\frac{1}{3}
\sum_{a,b,c} d_{abc}
 :\! \partial^\alpha \alpha^{(a)}(z) \partial^\beta \alpha^{(b)}(z)
     \partial^\gamma \alpha^{(c)}(z) \!:,
\nonumber\\
&&
W^{(4|\alpha,\beta,\gamma,\delta)}(z) =
\frac{1}{4}
\sum_{a,b,c,d,e} d_{abe} d_{cde}
 :\! \partial^\alpha \alpha^{(a)}(z) \partial^\beta \alpha^{(b)}(z)
     \partial^\gamma \alpha^{(c)}(z) \partial^\delta \alpha^{(d)}(z) \!:,
\qquad
\nonumber\\
&&
\hspace{60pt}\vdots
\eea
where $\alpha$, $\beta$, $\gamma$, $\delta$, \ldots 
run 0,1,2,\ldots and indicate the number of derivatives of $z$. 
However, some generators are not independent because of special properties of 
structure constants. As examples of such properties we mention 
\bea
&&
\sum_{b} d_{abb} = 0
\qquad\quad
\sum_{d,\,e,\,f} d_{ade} d_{bef} d_{cfd}
=
-\, \frac{d - 2}{12}\!\; d_{abc},
\nonumber\\
&&
\sum_e d_{\underline{ab}e} d_{\underline{cd}e}
=
\frac{6}{d + 2}
\sum_{e,\,f,\,g,\,h}
   d_{\underline{a}ef}
   d_{\underline{b}fg}
   d_{\underline{c}gh}
   d_{\underline{d}he}
=
\frac{1}{3}\, \delta_{\underline{ac}} \, \delta_{\underline{bd}},
\label{relations}
\eea
where the indices with underline is symmetrized. The implications of relations
like the ones listed in \rf{relations} will be discussed elsewhere \cite{to-come}.

Until now we have just treated the different operators 
$\alpha^{(a)}$ as indexed with a ``flavor''.   
The unbroken symmetry ($SO(2)$, $SO(3)$, $SO(5)$ and $SO(9)$) allows 
us to transform these 
flavors dimensions into each other. 
However, in the four cases 
we want space to be viewed as a 2, 3, 5 and 9 dimensional connected 
continuum, respectively.
Preferable, we  want to be able to talk about these spaces as topological spaces, 
e.g.\ 2, 3, 5 or 9 dimensional tori 
where the concept of neighborhoods or 
maybe even distances make some sense. 
This might happen dynamically via the cubic interaction, 
a possibility we find intriguing. 
Consider the  simplest situation:
$\tH_3(\mathbbm{R})$, and  symmetry breaking in the 8-direction. 
We have $d_{118}=d_{338} =  1/\sqrt{3}$ and $d_{888} = -1/\sqrt{3}$. 
Thus, according to one of the point of views presented above, 
space in the 8-direction will have no extension, 
but a 1-space and a 3-space can be glued together 
at a ``point" via 8-space. 
A set of such wormholes of 8-space, 
each of which has infinitesimal length and 
connects one point of 1-space and one point of 3-space, 
forms two-dimensional coordinates.
We can image such a ``knitting" taking place everywhere 
and in this sense the interaction via the 8-direction mode 
is what will create for us 
the genuine concept of a two-dimensional space for $\tH_3(\mathbbm{R})$.
Similar considerations apply for the higher dimensional spaces coming from 
$\tH_3(\mathbbm{C})$, $\tH_3(\mathbbm{H})$ and $\tH_3(\mathbbm{O})$. 
The knitting mechanism has the potential to form 
the space into a higher dimensional torus. 
Clearly this idea need to be substantiated by more explicit calculations. 

If we take the point of view that 
we allow ``negative" $L$ in the sense discussed above,
all directions now have an extension, also the 8-direction used for 
the ``knitting", and if we want such a ``knitting" picture 
to make sense we have create
large extended spaces and small spaces  of ``Planck size" and the   
8-direction have to be of Planck size. This can be done
by having positive and negative ``cosmological" terms 
($\mu$ and $\tilde{\mu}$ in the notation of eqs.\ \rf{yj1} and \rf{yj2}) 
and insisting that the scale $1/\sqrt{\mu}$ 
should be viewed as the Planck scale. 
Then, depending on the symmetry breaking 
mass matrix, a number of dimensions will be of Planck size, 
while the others will expand infinitely. 
Of course this picture is even more challenging than the picture 
where some flavors could not be associated with any spatial extension, 
since the flavors with macroscopic spatial extension acquire this extension 
within Planck time (see eq.\ \rf{yj2}). 
One needs a mechanism which slows down this expansion and 
in this context it is natural to think about Coleman's mechanism 
for lowering the cosmological constant \cite{coleman}. 
Again we clearly need explicit calculations to substantiate any claims, 
but contrary to Coleman's situation we actually have a model 
where questions of baby universe creation and annihilation can be 
answered by calculations, although going beyond perturbation theory 
when it comes to the creation and annihilation might be difficult 
(some calculation to all order exists 
in CDT string field theory \cite{sumgenus}).

As always, a model for the Big Bang and for the creation of the universe from 
nothing creates more questions than it answers. This is also the case for this 
model. However, it is an explicit model where hopefully explicit calculations 
can be performed, and it might be that some cosmological predictions do not 
require the full solution of the model 
and thus can be used to falsify the model
when compared to observations. Alternatively they might be encouraging and 
then provide further motivation for studying the model.  As an example we have 
very preliminary indications that the model can provide an explanation of dark 
energy which is not related to any ``bare" cosmological constant which 
might appear in the process of symmetry breaking of the $\Wtre$ algebra.
This will be published elsewhere, once we feel more confident of a number 
of other features of the model. 

\vspace{24pt} 

\noindent {\bf Acknowledgments.} 

JA and YW acknowledge support from the ERC-Advance grant 291092,
``Exploring the Quantum Universe'' (EQU).

%\vspace{1cm}

\end{document}